\documentstyle[12pt]{article}
\begin{document}

\begin{flushright}
NUB-TH-3138/96\\
CTP-TAMU-38/96\\
\end{flushright}

\begin{center}
{\bf Baryon Instability in SUSY Models}
\end{center}
 
\begin{center}
{Pran  Nath\\
Department of Physics, Northeastern University\\
Boston, MA  02115, USA}\\
{R. Arnowitt\\
Center for Theoretical Physics, Department of Physics\\
Texas A \& M University, College Station, TX77843, USA}
\end{center}

\begin{abstract}
 A review is given of nucleon instability in SUSY models.
The minimal SU(5) model is discussed in detail.
 \end{abstract}
 
{\bf 1. ~~Introduction}:

We begin by discussing proton instability in non-supersymmetric
grand unification. The simplest unified model that accomodates 
the electro-weak and the strong interactions is the SU(5) model\cite{gg}
and the instability
of the proton arises here from the lepto-quark exchange with mass $M_V$.
The dominant decay is the $e^+\pi^0$ mode and its lifetime can be 
written in the form\cite{gm}

\begin{equation}
\tau(p\rightarrow e^+\pi^0) \approx (\frac{M_V}{3.5 \times 10^{14}  
GeV})^4 10^{31\pm 1} yr
\end{equation}
The current experimental limit of this decay mode is\cite{pdg}  

\begin{equation}
\tau(p\rightarrow e^+\pi^0)>9\times  10^{32} yr, (90\% CL)
\end{equation}
In non-SUSY SU(5) the $e^+\pi^0$ mode has a partial lifetime of
$\tau(p\rightarrow e^+\pi^0)\leq 4\times 10^{29\pm 2}$ yr.Thus the 
non-SUSY SU(5) is ruled out because of the p-decay experimental limits.
It is expected that the Super Kamiokande will increase the sensitivity
of this mode to $1\times 10^{34}$\cite{totsuka}.
 That would imply that theoretically
the $e^+\pi^0$ mode would be observable if $M_V\leq 5\times 10^{15}$ GeV.
In supersymmetric grand unification  
current analyses based on unification of couplings constants already put 
a constraint on $M_G$ of about $10^{16}$\cite{deboer}.
Thus it seems not likely that 
the $e^+\pi^0$ mode would be observable in supersymmetry even in the 
next generation of proton decay experiments. Infact, reasonable estimates
indicate that $ \tau(p\rightarrow e^+\pi^0)> 1\times  10^{37\pm 2} yr$.

	In supersymmetric unification the dominant instablity of the 
proton arises via baryon number violating dimension five 
operators\cite{wein, acn,nac, hisano}.
In SUSY SU(5) operators of this type arise from the exchange of 
Higgs triplet fields and they have chiral structures LLLL and RRRR 
in the superpotential after the superheavy Higgs triplet field is 
eliminated. The main decay mode of the proton in these models is 
$\tau(p\rightarrow \bar \nu K)$.The current experimental limit for
this decay mode is\cite{pdg}
\begin{equation}
\tau(p\rightarrow \bar \nu K)> 1.0\times  10^{32} yr
\end{equation}
It is expected that Super Kamiokande will reach a sensitivity of 
 $2\times  10^{33}$ yr\cite{totsuka} while ICARUS will reach a sensitivity of
$5\times 10^{33}$ yr\cite{icarus}.
 Thus it is an interesting question to explore
to what extent the new generation of proton decay experiments will 
be able to test SUSY unified models.Actually we shall show that, unlike the
prediction for the $e^+\pi^0$ mode, it is not possible to make any
concrete predictions for the $\bar \nu K$ mode in SUSY models 
without inclusion of the low energy SUSY mass spectra which depends 
on the nature of supersymmetry breaking. Such an explicit supersymmetry
breaking mechanism is provided in supergravity grand unification\cite{can,bfs},
 but 
not in MSSM.Thus it is only in supergravity grand unification\cite{can}
 that one
can make detailed meaningful predictions of proton decay lifetimes.\\
{\bf 2. GUT Varieties}

Even within supersymmetric framework there are many possibilities
that may occur. The simplest of these is the minimal
SU(5) model. However, one can have extended gauge groups  such as
SU(3)$^3$, SO(10),...etc. and also string inspired models such as
SU(5)$\times$U(1)\cite{flipped}.There has been several works in the literature 
	 where there is a suppression of
 dimension five proton decay operators.There are a variety of ways in which
a  suppression of p-decay can occur\cite{flipped,BB,moh}. 
One possibility is that
matter is embedded in some unusual fashion in the basic particle
multiplets. Such a situation arises in the flipped SU(5)$\times U(1)$
model where one has an interchange $u\leftarrow\rightarrow d$ 
and  $e\leftarrow\rightarrow \nu$ relative to the usual embeddings.
 The other possibilty is the presence of 
some discrete symmetry which might forbid
the baryon number violating dimension five operators. In the following 
we discuss the  condition that would forbid such operators in
the general case. Let us assume that one has several Higgs triplets 
1,2,.,N 
that couple with the matter fields. We make a field redefinition 
so that the linear combination that couples with matter is labelled
$H_1,\bar H_1$ while the remaining Higgs triplet field have
no couplings. We may write their interactions
in the form 
\begin{equation}
\bar H_{1x}J^x + \bar K_x H_1^x+ \bar H_{ix} M_{ij}^{xy} H_{jy}
\end{equation}
where J and K are given by
\begin{equation}
J^x=\lambda^2 \bar M_y M^{xy}, ~K_x=\lambda^1 \epsilon_{xyzuv}M^{yz}M^{uv}
\end{equation}

Now the condition that the dimension five operators be suppressed is
given by\cite{testing}

\begin{equation}
(M^{-1})_{11}=0
\end{equation}
Of course satisfaction of the above condition would require either a 
finetuning or a discrete  symmetry.It is generally found that the 
imposition of discrete symmetries can lead to unwanted light higgs doublets
or light Higgs triplets\cite{coug}
 which would spoil the consistency of the 
unification of couplings with the LEP data.
  It is possible that string theory  may 
generate the desired discrete symmetries which suppress proton decay
without producing undesirable features alluded to above.However,
 more generally one can
expect proton decay to occur in both supergravity and string models.
The detailed nature of proton decay modes, their signatures and partial lifetimes
would depend on the specifics of the model.

 Another problem that 
surfaces in supersymmetric unified models is that of the doublet-triplet
splitting. That is one needs a mechanism that makes the Higgs triplets 
which mediate proton decay heavy and the Higgs doublet which 
generate electro-weak breaking light.Normally one simply finetunes the 
parameters to generate this splitting. Other possibilities consist of 
the so called
missing partner mechanism\cite{missing}, 
where one uses 75 , 50 and $\bar {50}$ 
representations in SU(5) instead of the usual 24 plet to break SU(5).
Here 50 contains  a ($\bar 3$, 1) and the $\bar{50}$ contains a 
(3,1) part in SU(3)$\times SU(2)$ decomposition but no (1,2)
pieces. Thus the Higgs triplets from the 50 and $\bar {50}$ will
match up with the Higgs triplet states from the 5 and $\bar 5$ when
the 75 plet develops a superheavy VEV, leaving the Higgs doublets 
light. More recently a mechanism has been dicussed in the 
literature which makes use of higher 
global symmetries such as SU(6) in the GUT sector which lead to
light Higgs as pseudo-Goldstone doublets\cite{pseudo} when the local SU(5)
symmetry breaks.However, there is as yet no complete
model which also gives acceptable pattern of masses to fermions for this 
mechanism.Several other mechanisms have also been discussed mostly in
  SO(10) frameworks\cite{dim, bb1,LM}
 and  make use of vacuum alignment,
 discrete symmetries etc to achieve the doublet-triplet splitting.
 In the following we shall assume that the doublet-triplet splitting  is
 resolved and discuss
the case of the minimal SU(5) model in detail .\\
{\bf 3. Nucleon Instability in Supergravity SU(5)}

	We discuss now the details of proton decay in the minimal 
SU(5) model to get an idea of the sizes of the lifetimes of the various
decay modes. The invariant potential of this model is given by
\begin{equation}
W_Y=-\frac{1}{8}f_{1ij}\epsilon_{uvwxy}H_1^uM_i^{vw}M_j^{xy}+
f_{2ij}\bar H_{2u}\bar M_{iv} M_j^{uv}
\end{equation}
Here the $M_{xi}, M^{xy}_i$ stand for the three generations(i=1,2,3)
of $\bar 5 , 10$ plet of 
quarks and leptons and $H_1,H_2$ are the $\bar 5,5 $ plet of Higgs
which give masses to the down and up quarks.
 After spontaneous breaking 
of the GUT group SU(5)$\rightarrow SU(3)\times SU(2)\times U(1)$ and 
integration over the heavy fields one has  the effective dimension
five operators with  baryon number violation given by
$\it L_5=L_5^L+L_5^R$ 
where\cite{acn,nac} 
\begin{eqnarray}
\it L^L_5= \frac{1}{M} \epsilon_{abc}(Pf_1^uV)_{ij}(f_2^d)_{kl}
( \tilde u_{Lbi}\tilde d_{Lcj}(\bar e^c_{Lk}(Vu_L)_{al}-
\nu^c_kd_{Lal})+...)+H.c.\nonumber\\
\it L^R_5= -\frac{1}{M} \epsilon_{abc}(V^{\dagger} f^u)_{ij}(PVf^d)_{kl}
(\bar e^c_{Ri}u_{Raj}\tilde u_{Rck}\tilde d_{Rbl}+...)+H.c.
\end{eqnarray}
Here the Yukawa couplings $f^u, f^d$ are related to the quark masses $m^u$
and $m^d$ as 
\begin{eqnarray}
m_i^u=f_i^u (sin2\theta_W/e)M_Z sin\beta\nonumber\\
m_i^d=f_i^d (sin2\theta_W/e)M_Z sin\beta
\end{eqnarray}
where $\theta_W$ is the Weak angle and $\beta$ is defined by 
$tan\beta=\frac{v_2}{v_1}$
where $v_2=<H_2^5>$ and $v_1=<H_1^5>$.  Further, V is the 
Kobayashi-Maskawa (KM ) matrix and P is a diagonal phase matrix with 
elements 
\begin{equation}
P_i=(e^{i\gamma_i}), ~\sum_i \gamma_i=0; ~i=1,2,3
\end{equation}

The dimension five operators must be dressed by the exhange of 
gluinos, charginos and neutralinos to produce dimension six 
operators which produce proton decay
Of all these exchanges the chargino exchange is the most dominant 
and is governed by the interaction

\begin{eqnarray}
L_{ui}^{\tilde W}=\frac{ig_2}{\sqrt 2}(cos\gamma_-\tilde W_1+
sin\gamma_-\tilde W_2)(V\gamma^0d_L)_i-ig_2(2cos\beta M_W)^{-1}\nonumber\\
(E cos\gamma_+\tilde W_1-
sin\gamma_+\tilde W_2)(Vm^d \gamma^0d_R)_i
\end{eqnarray}
where $\gamma_{\pm}$ are defined in the text preceding eq(30).
The dressing loop diagrams which convert dimension five into dimension
six operators also include squark and slepton exchanges. However, the
sfermion states that are exchanged are not pure L or R chiral states.
 As will be  discussed later, in Supergravity one has soft susy breaking
terms which mix the L and the R terms so that one has a (mass)$^2$ 
 matrix of the form 
\begin{equation}
\left({{{m_{\tilde{m}_{Rui}}^{2}}\atop{m_{i}^u(A_{ui}m_{o}-\mu
ctn\beta)}}{{m_{i}^u(A_{ui}m_{o}-\mu
ctn\beta)}\atop{m_{\tilde{t}_{Lui}}^{2}}}}\right)
\end{equation}
where $A,\mu$ are parameters which will be discussed in sec5.
The mass diagonal states are denoted by the scalar squark fields 
$\tilde u_{i(1,2)}$.These are related to the L and R chiral states
as 
\begin{equation}
\tilde u_{Ri}=cos\delta_{ui}\tilde u_{i1}+sin\delta_{ui}\tilde u_{i2},
~\tilde u_{Li}=-sin\delta_{ui}\tilde u_{i1}+cos\delta_{ui}\tilde u_{i2}
\end{equation}
where $\delta_{ui}$ is defined by 
\begin{equation}
sin2\delta_{ui}=-2 m_{ui}(A_{ui}m_0-\mu ctn\beta)/(\tilde m_{ui1}^2-\tilde
m_{ui2}^2)
\end{equation}
The chiral structure of the dimension six operators involves operators
of the type LLLL, LLRR, RRLL and RRRR. Of these it is the operators 
of the first type,i.e., LLLL which are the most dominant. In general
one finds many SUSY decay modes for the  proton,i.e., 
\begin{eqnarray}
\bar\nu_iK^+,\bar\nu_i\pi^+ ; i=e,\mu,\tau\nonumber\\
e^+K^0,\mu^+K^0,e^+\pi^0,\mu^+\pi^0,
e^+\eta,\mu^+\eta
\end{eqnarray}
The dependences of the branching ratios on quark mass factors and 
on KM matrix elements is shown in Table1.Also exhibited are the  
 enhancement factors, denoted by $y^{tk}_1 $ etc, from the third 
generation squark and slepton exchange contributions in the 
dressing loop diagrams.

\begin{center} \begin{tabular}{|c|c|c|c|}
\multicolumn{4}{c}{Table~1:~lepton + pseudoscalar decay modes of the proton
\cite{acn,nac} } \\
\hline
SUSY Mode & quark factors  & CKM factors & 3rd generation enhacement\\
\hline
$\bar \nu_eK$ &$m_d m_c$  &$V_{11}^{\dagger}V_{21}V_{22} $
& $(1+y_1^{tK}) $\\
\hline
$\bar \nu_\mu K $ &$m_s m_c$  &$V_{21}^{\dagger}V_{21}V_{22} $
& $(1+y_2^{tK})$\\
\hline
$\bar \nu_\tau K $ &$m_b m_c$  &$V_{31}^{\dagger}V_{21}V_{22} $
& $(1+y_3^{tK}) $\\
\hline
$\bar \nu_e \pi,\bar \nu_e \eta $ &$m_d m_c$  &$V_{11}^{\dagger}V_{21}^2 $
& $(1+y_1^{t\pi}) $\\
\hline
$\bar \nu_\mu \pi,\bar \nu_\mu \eta$ &$m_s m_c$  &$V_{21}^{\dagger}V_{21}^2 $
& $(1+y_2^{t\pi}) $\\
\hline
$\bar \nu_\tau \pi,\bar \nu_\tau\eta $ &$m_b m_c$  &$V_{31}^{\dagger}V_{21}^2 $
& $(1+y_3^{t\pi}) $\\
\hline
$eK $ &$m_d m_u$  &$V_{11}^{\dagger}V_{12} $
& $(1+y_e^{tK}) $\\
\hline
$\mu K $ &$m_s m_u$  & & $(1-V_{12}V^{\dagger}_{21}-y_{\mu}^{tK}) $\\
\hline
$e\pi, e\eta $ & $m_d m_u$ &  
&  $(1-V_{11}V^{\dagger}_{11}-y_e^{t\pi}) $  \\
\hline
$\mu \pi,\mu \eta $ &$m_s m_u$  & $V_{11}^{\dagger}V_{21}^{\dagger}$
&     $(1+y_{\mu}^{t\pi}) $\\
\hline
\end{tabular} 
\end{center}

The dependence of y factors, which contain the third generation contributions,
 on quark masses
and KM matrix elements is shown in Table2. 
The factors $R_e, R_{\mu},$ etc that enter in the evaluation of y in 
table2 are the dressing loop integrals and their explicit form 
is given in ref.\cite{nac}

\begin{center} \begin{tabular}{|c|c|}
\multicolumn{2}{c}{Table~2:  Third generation  
factors. } \\
\hline
y factor  & evaluation of y  \\
\hline
$y_1^{tK}$ &$ \frac{P_3}{P_2}\frac{m_t V_{31} V_{32}}{m_c V_{21}V_{22}}R_e$   \\
\hline
$y_2^{tK}$ &$ \frac{P_3}{P_2}\frac{m_t V_{31} V_{32}}{m_c V_{21}V_{22}}R_{\mu}
$\\
\hline
$y_3^{tK}$ &$ \frac{P_3}{P_2}\frac{m_t V_{31} V_{32}}{m_c V_{21}V_{22}}R_{\tau}
$\\
\hline
$y_1^{t\pi}$ &$ \frac{P_3}{P_2}\frac{m_t V_{31}^2}{m_c V_{21}^2}R_e$   \\
\hline
$y_2^{t\pi}$ &$ \frac{P_3}{P_2}\frac{m_t V_{31}^2}{m_c V_{21}^2}R_{\mu}$   \\
\hline
$y_3^{t\pi}$ &$ \frac{P_3}{P_2}\frac{m_t V_{31}^2}{m_c V_{21}^2}R_{\tau}$   \\
\hline
$y_e^{tK}$ &$ \frac{P_3}{P_1}\frac{m_t V_{32} V_{33}V_{31}^{\dagger}}
{m_u V_{12}}R_e'$   \\
\hline
$y_{\mu}^{tK}$ &$ \frac{P_3}{P_1}\frac{m_t V_{21}^{\dagger}V_{32} 
V_{33}V_{31}^{\dagger}}
{m_u}R_{\mu}'$   \\
\hline
$y_e^{t\pi}$ &$ \frac{P_3}{P_1}\frac{m_t V_{31} V_{33}V_{31}^{\dagger}
V_{11}^{\dagger}}
{m_u}R_e^{''}$   \\
\hline
$y_{\mu}^{t\pi}$ &$ \frac{P_3}{P_1}\frac{m_t V_{33} V_{31}^{\dagger}}
{m_u V_{11}}R_{\mu}^{''}$   \\
\hline
\end{tabular} 
\end{center}

From tables 1 and 2 one finds that 
there is a hierarchy in the partial decay branching ratios of these
modes which can be read off from the quark mass factors and the
KM matrix elements. In making order of magnitude 
 estimates for lifetimes it is 
useful to keep in mind that 
\begin{equation}
muV_{11}:m_cV_{21}:m_t V_{31}\approx 1:50:500
\end{equation}
One can then roughly order the partial decay branching ratios for the
various modes listed in table1 as follows 
\begin{eqnarray}
BR(\mu K)>>BR(\mu K),BR(\mu\pi)>>BR(\mu \pi)\nonumber\\
BR(\bar \nu K)>BR(\bar \nu \pi)> BR(\it l K)> BR(\it l \pi)\nonumber\\
BR(\bar \nu_{\mu}K)>BR(\bar \nu_{\tau} K)> BR(\bar \nu_e K),
~BR(\bar \nu_{\mu}\pi)>BR(\bar \nu_{\tau}\pi)> BR(\bar \nu_e\pi)
\end{eqnarray}
One finds from the above that the  most dominant decay modes of the 
proton are the $\bar \nu K$ modes.The dimension six operators which
govern these are given by\cite{acn,nac} 
\begin{eqnarray}
L_6(N\rightarrow \bar\nu_i K) =[(\alpha_2)^2 (2MM_W^2sin2\beta)^{-1}
P_2m_cm_i^d V_{i1}^{\dagger}V_{21} V_{22}]
[F(\tilde c,\tilde d_i, \tilde W)
+F(\tilde c,\nonumber\\
\tilde d_i, \tilde W)]
+([1+y_i^{tK}+(y_{\tilde g}+y_{tilde Z})\delta_{i2}+\Delta_i^K]\alpha_i^L
+[1+y_i^{tK}-(y_{\tilde g}-\nonumber\\
y_{\tilde Z})\delta_{i2}+\Delta_i^K]\beta_i^L
+(y_1(R)\alpha_3^R+y_2^{(R)}\beta_3^R)\delta_{i3})
\end{eqnarray}
In the above $\alpha_i^{L,R},\beta_I^{L,R}$ are defined by
\begin{equation}
\alpha_i^L=\epsilon_{abc}(d_{aL}\gamma^0u_{bL})(s_{cL}
\gamma^0\nu_{iL})
\end{equation}
and
$\alpha_i^R=\alpha_i^L$$(d_L,u_L\rightarrow d_R,u_R)$, and $~\beta_i^{L,R}=
\alpha_i^{L,R}(d\leftarrow\rightarrow s)$.
Further $y_i^{tk}$ gives the dominant contribution from the third generation
and is defined by\cite{acn,nac} 
\begin{equation}
y_i^{tK}=\frac{P_2}{P_3}(\frac{m_sV_{31}V_{32}}{m_c V_{21}V_{22}})
(\frac{F(\tilde t,\tilde d_i,\tilde W)+F(\tilde t,\tilde e_i,\tilde W)}
{F(\tilde c,\tilde d_i,\tilde W)+F(\tilde c,\tilde e_i,\tilde W)})
\end{equation}
where the functions F are dressing loop integrals and would be defined
explicitly below. The remaining contributions represented by $\Delta_i^K$,
$y_{\tilde g}$ ( from gluino exchange) and $y_{\tilde Z}$(from neutralino
exchange) are all relatively small. 

	The decay branching ratios of the p into the $\bar \nu_i K$
modes are given by the relation 
\begin{equation}
\Gamma(p\rightarrow \bar\nu_iK^+)=(\frac {\beta_p}{M_{H_3}})^2|A|^2
|B_i|C
\end{equation}
where $\beta_p$ is the three quark - vacuum matrix element of the
proton and is   defined by 
\begin{equation}
\beta_p U_L^{\gamma}=\epsilon_{abc}\epsilon_{\alpha \beta} <0|d_{aL}^{\alpha}
u_{bL}^{\beta}u_{cL}^{\gamma}|p>
\end{equation}
The most recent evaluation of $\beta_p$ is from lattice gauge calculations
\cite{gavela} and is
\begin{equation}
\beta_p=(5.6\pm 0.5)\times  10^{-3} GeV^3
\end{equation}
The factors A and $B_i$ of eq(21) are defined by 

\begin{equation}
A=\frac{\alpha_2^2}{2M_W^2}m_s m_c V_{21}^{\dagger} V_{21}A_L A_S
\end{equation}
\begin{equation}
B_i= \frac{1}{sin2\beta}\frac{m_i^d V_{i1}^{\dagger}}{m_sV_{21}^{\dagger}} 
[P_2 B_{2i}+\frac{m_tV_{31}V_{32}}{m_cV_{21}V_{22}} P_3B_{3i}]
\end{equation}

\begin{equation}
B_{ji}=F(\tilde u_i,\tilde d_j,\tilde W)+(\tilde d_j\rightarrow \tilde e_j)
\end{equation}
where 
\begin{eqnarray}
F(\tilde u_i,\tilde d_j,\tilde W)=[E cos\gamma_-sin\gamma_+\tilde f(\tilde
u_i,\tilde d_j, \tilde W_1)
+cos\gamma_+sin\gamma_-\tilde f(\tilde
u_i,\tilde d_j, \tilde W_1)]\nonumber\\
-\frac{1}{2} \frac{\delta_{i3}m_i^u sin2\delta_{ui}}{\sqrt 2 M_W sin\beta}
[E sin\gamma_-sin\gamma_+\tilde f(\tilde
u_{i1},\tilde d_j, \tilde W_1)
-cos\gamma_-cos\gamma_+\tilde f(\tilde
u_{i1},\tilde d_j, \tilde W_2)\nonumber\\
 - (\tilde u_{i1}\rightarrow \tilde u_{i2})]
\end{eqnarray}
In the above $\tilde f$ is given by 
\begin{equation}
\tilde f(\tilde u_i,\tilde d_j, \tilde W_k)=sin^2\delta_{ui}
\tilde f(\tilde u_{i1},\tilde d_j, \tilde W_k)
+cos^2\delta_{ui}
\tilde f(\tilde u_{i2},\tilde d_j, \tilde W_k) 
\end{equation}
where 
\begin{equation}
f(a,b,c)=\frac{m_c}{m_b^2-m_c^2}[\frac{m_b^2}{m_a^2-m_b^2}ln(\frac{m_a^2}
{m_b^2})-(m_a\rightarrow m_c)]
\end{equation}
and $ \gamma_{\pm}=\beta_+\pm\beta_- $ where 
\begin{equation}
sin2\beta_{\pm}=\frac{(\mu\pm \tilde m_2)}{[4\nu_{\pm}^2
+(\mu\pm \tilde m_2)^2]^{1/2}}
\end{equation}
and 
\begin{equation}
\sqrt 2 \nu_{\pm}=M_W(sin\beta\pm cos\beta)
\end{equation}
\begin{equation}
sin2\delta_{u3}=-\frac{-2(A_t+\mu ctn\beta)m_t}{m_{\tilde t_1}^2-
m_{\tilde t_2}^2}
\end{equation}
\begin{eqnarray}
E=1~, sin2\beta>\mu\tilde m_2/M_W^2 \nonumber\\
~~~=-1,sin2\beta<\mu\tilde m_2/M_W^2  
\end{eqnarray}
Finally C that enters eq(25) is a current algebra factor and is given by  
\begin{equation}
C=\frac{m_N}{32\pi f_{\pi}^2} [(1+\frac {m_N(D+F)}{m_B})
(1-\frac{m_K^2}{m_N^2})]^2
\end{equation}
where  the chiral Lagrangian factors $f_{\pi}, D,F, ..$ etc 
that enter the above equation have the numerical values:
$ f_{\pi}=139$~MeV,D=0.76,F=0.48,$m_N$=938 ~MeV, ~$m_K$=495 ~MeV, 
and  ~$m_B$=1154.\\
{\bf 4.~Vector ~Meson ~Decay ~Modes ~of ~the ~Proton}

The same baryon number violating dimension six quark operators that 
lead to the decay of the proton into lepton and pseudoscalar modes
also lead to decay modes with lepton and vector mesons\cite{yuan}.
 Although the
vector mesons are considerably heavier that their corresponding pseudoscalar
counterparts, decay modes invoving $\rho,K^*,\omega$ are still allowed.
We list these below
\begin{eqnarray}
\bar\nu_iK^*,\bar\nu_i\rho,\bar\nu_i\omega ; i=e,\mu,\tau\nonumber\\
e K^*,\mu K^*,e\rho,\mu\rho,e\omega,\mu\omega
\end{eqnarray}
The 
quark, KM and third generation enhancement factors for the allowed vector
meson decay modes is exhibited in table3. The branching ratios for the
vector meson decay modes are typically smaller than the corresponding 
pseudo-scalar decay modes.

\begin{center} \begin{tabular}{|c|c|c|c|}
\multicolumn{4}{c}{Table~3: lepton + vector meson decay modes of the proton } \\
\hline
SUSY Mode & quark factors  & CKM factors & 3rd generation enhacement\\
\hline
$\bar \nu_eK^* $ &$m_d m_c$  &$V_{11}^{\dagger}V_{21}V_{22} $
& $(1+y_1^{tK}) $\\
\hline
$\bar \nu_\mu K^*$ &$m_s m_c$  &$V_{21}^{\dagger}V_{21}V_{22} $
& $(1+y_2^{tK}) $\\
\hline
$\bar \nu_\tau K^*$ &$m_b m_c$  &$V_{31}^{\dagger}V_{21}V_{22} $
& $(1+y_3^{tK}) $\\
\hline
$\bar \nu_e \rho,\bar \nu_e\omega $ &$m_d m_c$  &$V_{11}^{\dagger}V_{21}^2 $
& $(1+y_1^{t\pi}) $\\
\hline
$\bar \nu_\mu \rho,\bar \nu_\mu\omega$ &$m_s m_c$  &$V_{21}^{\dagger}V_{21}^2 $
& $(1+y_2^{t\pi}) $\\
\hline
$\bar \nu_\tau \rho,\bar \nu_\tau\omega$ &$m_b m_c$  &$V_{31}^{\dagger}V_{21}^2 $
& $(1+y_3^{t\pi}) $\\
\hline
$ e K^*$ &$m_d m_u$  &$V_{11}^{\dagger}V_{12} $
& $(1+y_e^{tK}) $\\
\hline
$\mu K^*$ &$m_s m_u$  &
& $(1-V_{12}V^{\dagger}_{21}-y_{\mu}^{tK}) $\\
\hline
$e\rho,e\omega$ &$m_d m_u$ &  
&  $(1-V_{11}V^{\dagger}_{11}-y_e^{t\pi}) $  \\
\hline
$\mu \rho,\mu \omega$ &$m_s m_u$  & $V_{11}^{\dagger}V_{21}^{\dagger}$
&     $(1+y_{\mu}^{t\pi}) $\\
\hline
\end{tabular} 
\end{center}

{\bf 5. Details of Analysis in Supergravity Unification}

	Next we discuss the details of the proton decay analysis
in supergravity unification. 
As already indicated the low energy SUSY spectrum plays a crucial role
in  proton decay
lifetime.In fact the spectrum that enters consists of 12 squark states,
9 slepton states, 4 neutralino states, 2 chargino states, and the gluino.
There are thus 28 different mass parameters alone. In globally 
supersymmetric grand unification one has no way to meaningfully
control these parameters and thus detailed predictions of p decay lifetimes
in globally supersymmetric theories cannot be made. In supergravity
unified models one has a well defined procedure of breaking supersymmetry
via the hidden sector and the minimal supergravity unification contains 
only 4 SUSY parameters in terms of which all the SUSY masses can be predicted.
Thus supergravity unification is very predictive.
 We give below a brief review of the basic 
elements of supergravity grand 
unification. These are: (1) supersymmetry breaks in the hidden sector by
a superhiggs phenomenon and the breaking of supersymmetry is 
communicated gravitationally to the physical sector; (2) the superhiggs
coupling are assumed not to depend on the generation index, and (3) 
 one assumes the spectrum to be the MSSM spectrum below the GUT scale.
After the breaking of supersymmetry and of the gauge group one can 
integrate over the superhiggs fields  and the heavy fields and the following
supersymmetry breaking potential in the low energy domain results
\cite{can,bfs}: 
\begin{equation}
V_{SB}=m_0^2 z_az_a^{\dagger}+(A_0W^(3)+B_0W^(2) +h.c.) 
\end{equation}

where $W^{(2)}, W^{(3)}$ are the bilinear and trilear parts of the 
superpotential.
	 There is also a gaugino mass term 
$\it L^{\lambda}_{mass}=-m_{1/2}\bar \lambda^{\alpha}\lambda^{\alpha}$.
At this stage the theory has five SUSY  parameters
$m_0, m_{1/2}, A_0,B_0, and ~\mu_0$.
Here $\mu_0$ is  the Higgs mixing term which along with the other low 
energy quark-lepton-
Higgs interactions is given by 
\begin{equation}
W=\mu_0H_1H_2 +[\lambda_{ij}^{(u)}q_iH_2u^C_j+
\lambda_{ij}^{(d)}d_iH_1d^C_j+\lambda_{ij}^{(e)}l_iH_1d^C_j]
\end{equation}
In the above  $H_1$ is the light Higgs doublet which gives mass
to  down quark and leptons and $H_2$ give mass to the up quark.
The number of SUSY parameters  can be reduced after radiative breaking
of the electro-weak symmetry. The radiative electro-weak symmetry 
breaking is governed by the potential
\begin{eqnarray}
V_H=m_1^2(t)|H_1|^2+m_2^2(t)|H_2|^2-m_3^2(t)(H_1 H_2 +h.c.) \nonumber\\ 
+\frac{1}{8} (g^2+g_y^2)(|H_1|^2-|H_2|^2)^2 +\Delta V_1
\end{eqnarray}
where  
 $\Delta V_1$ is the correction from one loop, and 
$m_1^2(t)$ etc  
 are the running parameters and  satisfy the boundary conditions 
$m_i^2(0)=m^2_0+\mu^2_0; i=1,2$,
$m_3^2(0)=-B_0\mu_0$,
$\alpha_2(0)=\alpha_G=(5/3)\alpha_Y(0)$.
The breaking of the elctroweak symmetry is accomplished by the 
relations
$\frac{1}{2}M_Z^2=(\mu_1^2-\mu_2^2 tan^2\beta)/(tan^2\beta-1)$
and  
$~sin2\beta=(2m^2_3)/(\mu_1^2+\mu_2^2)$, 
where $\mu_i^2=m_i^2+\Sigma_i$ and 
$\Sigma_i$ is one loop correction from $\Delta V_1$.Using the above relations
one can reduce the low energy SUSY parameters to the following:
\begin{equation}
m_0, m_{1/2}, A_0, tan\beta
\end{equation}
Another result that emerges from radiative breaking of the electro-weak 
symmetry is that of scaling.One finds that over most of the parameter
space of the theory $\mu^2>>M_Z^2$ which gives\cite{scaling1,scaling2}
\begin{eqnarray}
m_{\tilde W_1}\sim \frac{1}{3} m_{\tilde g}~( \mu<0) ;
m_{\tilde W_1}\sim \frac{1}{4} m_{\tilde g}~(\mu>0)\nonumber\\
2 m_{\tilde Z_1}\sim m_{\tilde W_1}\sim m_{\tilde Z_2};
m_{\tilde Z_3}\sim m_{\tilde Z_4}\sim m_{\tilde W_2} >> 
m_{\tilde Z_1}\nonumber\\
m_H^0\sim m_A\sim m_{H^{\pm}}>>m_h
\end{eqnarray}
  Corrections to the above are typically small  O(1/$\mu$) over most
of the parameter space.

	We discuss now the effects of the top quark which play an 
important role in limiting the parameter space of the model.
Contraints from the top quark arise because there is a Landau pole
in the top quark Yukawa coupling, i.e., 
$Y_0=Y_t/(E(t)D_0)$ where $D_0=1-6Y_t F(t)/E(t)$,
$Y_t=\lambda_t^2/4\pi$, $\lambda_t(Q)$ is the top-quark Yukawa coupling
and is defined by $m_t=<H_2>\lambda_t(m_t)$, and the functions E(t)
and F(t) are as defined in ref\cite{ilm}.We see from the above that 
the top Yukawa has a Landau pole which appears at 
\begin{equation}
m_t^f=(8\pi/\alpha_2(t))^{1/2}(Y_t^f(t))^{1/2}M_Zcos\theta_W sin\beta
\end{equation}
where $\theta_W$ is the weak mixing angle.For some typical values of
$\alpha_G$ and
$M_G$ one has 
$m_t^f\sim 200 sin\beta$.
Now it is found that the same Landau singularity also surfaces in the 
other SUSY parameters because of the coupled nature of the renormalization
group equations.Thus, for example, the trilinear soft SUSY parameter 
develops  a Landau singularity:
$A_0=A_R/D_0 + A_0(nonpole)$, and  $A_R=A_t-0.6 m_{\tilde g}$,  
where $A_0$ is the value of $A_t$ at the GUT scale.  A similar analysis 
shows that $\mu^2$ and thus the stop masses become singular.Specifically
one finds that
$m_{\tilde t_1}^2=-2x/D_0+ m_{\tilde t_1}^2(NP)$ and 
$m_{\tilde t_2}^2=-x/D_0+ m_{\tilde t_2}^2(NP)$
where x=$Y_tA_R^2F/E$. We note that the Landau pole contribution 
 is negative definite and thus 
drives the stops towards their tachyonic limit.
Especially  the Landau pole contributions to $\tilde t_1$ are rather
large and so its transition to the   
tachyonic limit is very rapid. Thus the condition that there be no tachyons
puts a strong  limit  on the parameter space. One finds that the allowed 
values of  $A_t$ lie in the range $ -0.5<A_t<5.5$.\\
 
{\bf 6. Discussion of Results}

Figure 1a gives the maximum lifetime of the $p\rightarrow \nu K^+$
mode for$\mu <0 $ as function of $m_0$ when all other parameters 
are varied over the allowed  parameters space consistent with 
radiative breaking of the electro-weak symmetry and with the
inclusion of the LEP1.4 constraints. The solid curve gives
the maximum without the imposition of the cosmological relic density 
constraint while the dashed curve includes the relic density
constraint. \\
The solid horizontal line is the current experimental 
lower limit for this mode from IMB and Kamiokande.
We see that
the analysis shows that there exists a cosiderable part of the
parameter space not yet explored  by the current experiment, which
will be accessible to SuperKamiokande and ICARUS. Figure 1b gives the
same analysis when $\mu >0$.Comparison of figs 1a and 1b shows that
the current experiment excludes a somewhat larger region of the 
parameter space in $m_0$ for $\mu >0$ than for $\mu <0$.Thus for
$\mu >0$ one eliminates the region $\mu < 400 GeV$ while for
$\mu <0$ only the values $m_0 <300 GeV$ are eliminated. Figure 1c
gives the plot of the maximum lifetime for the $p\rightarrow \bar \nu K$
mode as a function of gluino mass for the case $\mu <0$ corresponding
to Fig 1a. We see that regions of the parameter space with lifetimes 
above the current limits lie below approximately 400 GeV when the dark matter
constraint is imposed. Figure 1d is similar to Fig 1c except  that 
$\mu > 0$.\\

{\bf 7.Conclusion}

In the above we have given a brief review of nucleon instability in
supersymmetric unified theories. We have pointed out that no concrete
predictions on proton decay lifetimes are possible unless 
the nature of the low energy SUSY mass spectrum which enters in the
dressing loop diagrams is assumed. Thus no concrete predictions on proton
lifetime  can be made in globally supersymmetric grand unified
theories. In contrast one can make predictions in supergravity 
unification since the SUSY breaking spectrum of the theory is 
characterised by four parameters.Further since  there are 32 
supersymmetric particles one has 28 predictions in the model, and
thus supergravity grandunification is very predictive. An updated
analysis of p-decay in the minimal SU(5) model was given including the
constraints of LEP1.4.It is found that 
there exits a significant part of the parameter space which is
not yet explored by the current proton lifetime limits on 
$p\rightarrow \bar \nu K$  from IMB and Kamiokande but which 
 would become accessible to SuperKamiokande and ICARUS experiments.
Finally we note that 
the minimal model can correctly accomodate the $b/\tau$ mass 
ratio\cite{bbo}.
However, it does not predict other quark lepton 
mass ratios correctly and non-minimal extensions are needed for this purpose.
These non-minimal extensions also affect the proton lifetime predictions.

{\bf Acknowledgements}:
This research was supported in part by NSF grant numbers 
PHY-19306906  and  PHY-9411543.\\


{\bf References}
\begin{enumerate}

\bibitem{gg} H.Georgi and S.L.Glashow, Phys.Rev.Lett.{\bf 32},
438(1974).
\bibitem{gm}M.Goldhaber and W.J. Marciano, Comm.Nucl.Part.Phys.{\bf 16},
23(1986); P.Langacker and N.Polonsky, Phys.Rev.{\bf D47},4028(1993).

\bibitem{pdg} Particle Data Group, Phys.Rev. {\bf D50},1173(1994).

\bibitem{totsuka} Y.Totsuka, Proc. XXIV Conf. on High Energy Physics,
Munich, 1988,Eds. R.Kotthaus and J.H. Kuhn (Springer Verlag, Berlin, 
Heidelberg,1989).

\bibitem{deboer} W.de Boer, Prog.Part. Nucl.Phys.{\bf 33},201(1994).

\bibitem{wein} S.Weinberg,~Phys.Rev.{\bf D26},287(1982); 
N.Sakai and T.Yanagida, Nucl.Ph\\
ys.{\bf B197},
533(1982); S.Dimopoulos, S.Raby  and F.Wilcek, Phys.Lett.\\
 {\bf 112B}, 133(1982);
J.Ellis, D.V.Nanopoulos and S.Rudaz, Nucl.Phys.\\
{\bf  B202},43(1982);
B.A.Campbell, J.Ellis and D.V.Nanopoulos,\\
 Phys.Lett.{\bf 141B},299(1984);
S.Chadha, G.D.Coughlan, M.Daniel\\
 and G.G.Ross, Phys.Lett.{\bf 149B},47(1984).

\bibitem{acn} R.Arnowitt, A.H.Chamseddine and P.Nath, Phys.Lett.
{\bf 156B},215(1985).
\bibitem{nac}
P.Nath, R.Arnowitt and A.H.Chamseddine, Phys.Rev.{\bf 32D},2348(1985).

\bibitem{hisano} J.Hisano, H.Murayama and T. Yanagida, Nucl.Phys.
{\bf B402},46(1993).

\bibitem{icarus} ICARUS Detector Group, Int. Symposium on Neutrino 
Astrophsyics, Takayama. 1992.

\bibitem{can}
A.H. Chamseddine, R. Arnowitt and P. Nath, Phys. Rev. Lett {\bf 29}.
970 (1982);P.Nath,Arnowitt and A.H.Chamseddine ,
``Applied N=1 Supergravity" (World Scientific,
Singapore, 1984);
 H.P. Nilles, Phys. Rep. {\bf 110}, 1 (1984); R. Arnowitt and
P. Nath, Proc of VII J.A. Swieca Summer School (World Scientific, Singapore
1994).
\bibitem{bfs}
R.Barbieri, S.Ferrara and C.A.Savoy, Phys. Lett.{\bf B119}, 343(1982);
L.Hall, J.Lykken and  S.Weinberg, Phys. Rev. {\bf D27}, 2359(1983);
P.Nath, R.Arnowitt and A.H.Chamseddine, Nucl. Phys. {\bf B227}, 121(1983).

\bibitem{flipped} I.Antoniadis, J.Ellis, J.S.Hagelin and D.V.Nanopoulos,
Phys.Lett.{\bf B231},65\\
(1987); ibid, {\bf B205}, 459(1988).

\bibitem{BB} K.S.Babu and S.M. Barr,Phys.Rev. {\bf D48}, 5354(1998).

\bibitem{moh} R.N.Mohapatra, UMD-PP-96-59(1996)/hep-ph/9601203.


\bibitem{testing} R.Arnowitt and P.Nath, Phys.Rev. {\bf D49}, 1479(1994).


\bibitem{coug} C.D. Coughlan, G.G.Ross, R.Holman, P.Ramond, M.Ruiz-
Altaba and J.W.F. Valle, Phsy.Lett.{\bf 158B},401(1985).

\bibitem{missing}B.Grinstein, Nucl.Phys.{\bf B206},387(1982);H.Georgi,
Phys.Lett.{\bf B115},380\\
(1982).

\bibitem{pseudo} K.Inoue, A.Kakuto and T.Tankano, Prog.Theor. Phys.{\bf 75},
664(1986); A.Anselm and A.Johansen, Phys.Lett.{\bf B200},331(1988);
A.Anselm, Sov. Phys.JETP{\bf 67},663(1988); R.Barbieri, G.Dvali and 
A.Strumia, Nucl. Phys.\\
{\bf B391},487(1993); Z.Bereziani and G.Dvali, 
Sov.Phys. Lebedev Inst. Report 5,55(1989);Z.Bereziani, C.Csaki and L.
Randall, Nucl.Phys.{\bf B44},61\\
(1995).

\bibitem{dim} S.Dimopoulos and F.Wilczek,  Report No.NSF-ITP-82-07(1981)
(unpublished).
\bibitem{bb1}K.S.Babu and S.M.Barr, Phys.Rev.{\bf D50},3529(1994).
\bibitem{LM} D.Lee and R.N.Mohapatra, Phys.Rev.{\bf D51} (1995).

\bibitem{gavela} M.B.Gavela et al, Nucl.Phys.{\bf B312},269(1989).


\bibitem{yuan} T.C.Yuan, Phys.Rev.{\bf D33},1894(1986). 

\bibitem{scaling1}
R. Arnowitt and P. Nath, Phys. Rev. Lett. {\bf 69}, 725 (1992).
\bibitem{scaling2}
 P. Nath and R.
Arnowitt, Phys. Lett. {\bf B289}, 368 (1992).


\bibitem{ilm} L.Ibanez, C.Lopez, and C.Munos, Nucl. Phys.{\bf B256},
218(1985).

\bibitem{nwa}
P. Nath, J. Wu and R. Arnowitt, Phys.Rev.{\bf D52},4169(1995).

\bibitem{bbo}
V.Barger, M.S.Berger, and P.Ohman, Phys.Lett. {\bf B314},351(1993). 

\end{enumerate}
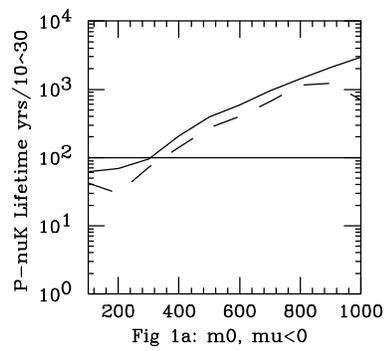
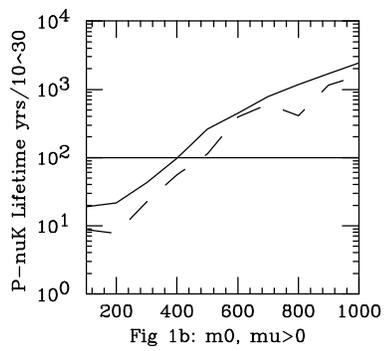
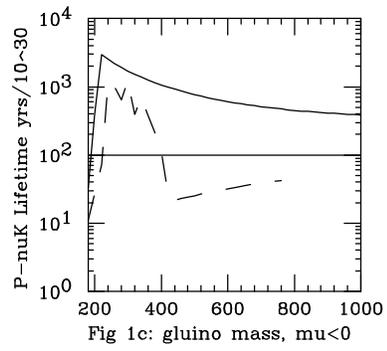
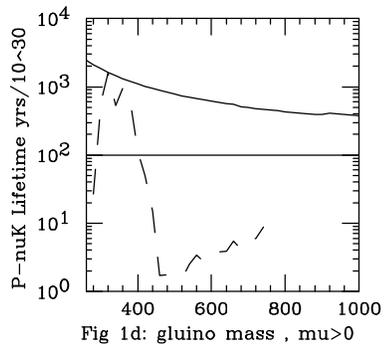
\end{document}